# Characterization and Modeling of an Electro-thermal MEMS Structure

P. G. Szabó, V. Székely
Budapest University of Technology and Economics (BME), Department of Electron Devices
H-1111 Budapest, Goldman György Tér 3., Hungary

*Abstract-* **Thermal functional circuits are an interesting and perspectivic group of the MEMS elements. A practical realization is called Quadratic Transfer Characteristic (QTC) element which driving principle is the Seebeck-effect. In this paper we present the analyses of a QTC element from different perspectives. To check the real behaviour of the device, we measured a few, secondary properties of the structure which correspond to special behaviour because these properties can not be easily derived from the main characteristics.**

## I. Introduction

Micro-electromechanical systems are tiny microdevices, which dimensions are in the range of 1 μm to several millimetres. An interesting group of these devices works in the electro-thermal domain. Several actuator and sensor applications are being developed where the parameters of the thermally induced structure changes, resulting some sort of an observable effect. An example for this issue comes up, if an electronically generated thermal energy is being converted back to electricity based on the Seebeck-effect. Utilizing this signal, special microsystems can be developed with several functions. A group of them is called thermal functional circuits, where different input-output characteristics can be achieved by varying the layout of thermopiles and heating elements. With these variants thermal multiplier, divider, RMS meter, cross-correlation meter and even modulator and demodulator can be developed.

The advantage of these circuits is that they can be fed on the temperature difference, generated by the losses of heat dissipating elements and therefore they do not load the drive circuit electrically. Their maximum speed is small compared to electrical circuits and the output amplitude is relatively low which make them very sensible to both on-chip and external noise.

In this paper we present the analyses of the basic component of the thermal functional circuits, named QTC element in [1]. In the first part, the mapping and functional verification takes place where the calculation of thermal parameters is discussed as well. The second section is about the determination of the thermal dependency of the system and in the last segment the examination of the upper harmonics in the response signal is described. Here, the electrical equivalent model of the MEMS is set up which is suitable to be used in SPICE like circuit simulation programs.

## II. Mapping and Function Verification

### A. Mapping and calculation of the thermal parameters

In the first chapter we present the elementary analysis of the electro-thermal test-MEMS. The testchip was made in TIMA Laboratory and it contains more, basically thermal microsystems. The thermal functional system was fabricated by a combination of bulk micromachining and CMOS technology. The investigated circuit itself was realized on a cantilever which trails over an etched trench. Besides the connection wires it contains a heating resistor which is responsible for creating the temperature difference in the cantilever and serially connected thermocouples to produce greater output voltage amplitude. Figure 1 shows the scanning electron microscopic (SEM) image of the implemented system. The heating resistor is at the left end of the cantilever with 3 μm width and 86 μm length. 4.3 μm right of it the hot points of the thermopiles were placed. 12 pieces of polysilicon-aluminum thermocouple were fabricated with 1.4 μm width and 213.8 μm length. They are connected to each other at the right end of the half-bridge where their cold points are.

Knowing the geometrical sizes and the thermal properties of the different layers form [2], the thermal parameters of the system can be calculated. First, the heat resistance and the heat capacity were calculated based on the area compensated resistive and capacitance parameters.

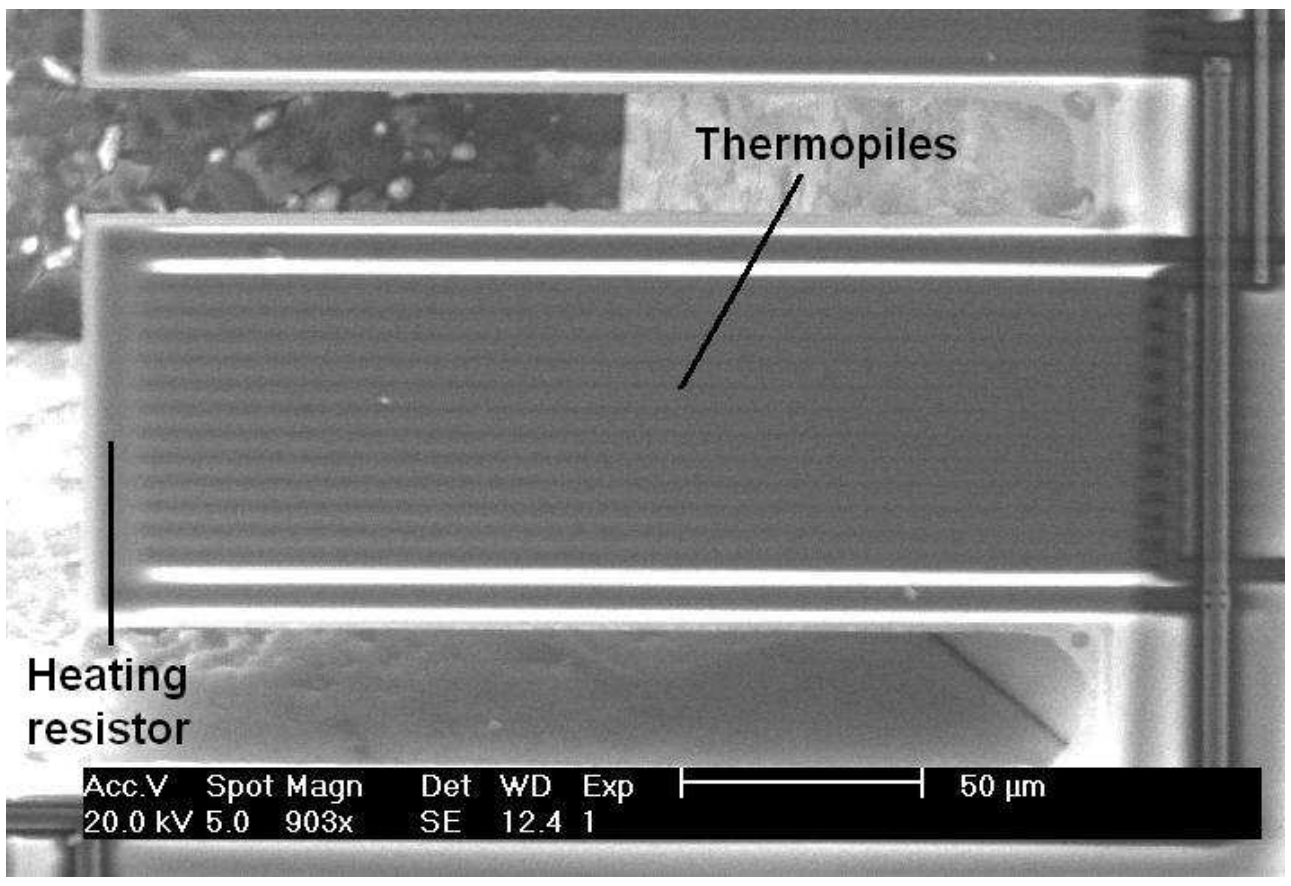

Figure 1. SEM image of the structure

$$R_{th} = \frac{\sum d_i}{\sum \lambda_i \cdot d_i} \cdot \frac{L}{A} = 56570 \frac{K}{W} \quad (1)$$

where $d_i$ is the thickness and $\lambda_i$ is the heat conductance of layer $i$, while $L$ is the length and $A$ is the cross-sectional area of the cantilever where the heat flow. The heat capacitance was calculated in similar way.

$$C_{th} = \frac{\sum C_i \cdot d_i}{\sum d_i} \cdot L \cdot A = 11.6 \cdot 10^{-8} \frac{W \cdot s}{K} \quad (2)$$

where $C_i$ is the heat capacity of layer $i$. As for this parameters are considered as the elements of a distributed R-C network. Their resultant impedance was calculated as follows [3] [4]:

$$Z(s) = \frac{R_{th}}{\sqrt{s \cdot R_{th} \cdot C_{th}}} \cdot \frac{\sinh \sqrt{s \cdot R_{th} \cdot C_{th}} \cdot \frac{x}{L}}{\cosh \sqrt{s \cdot R_{th} \cdot C_{th}}} \quad (3)$$

where $L$ is the length of the cantilever and $x$ is a smaller than $L$ arbitrary point.

Instead of a distributed network it is easier to cope with a lumped network, therefore we determined the Foster equivalent circuit of the cantilever and transformed it to Cauer R-C ladder. First, the poles and time-constants were needed to be calculated. Since the higher order poles lie relatively far, it is enough to cope with the first two poles which result in a two stage R-C ladder.

$$p_n = -\frac{\pi^2}{4} \cdot \frac{(2 \cdot n - 1)^2}{R_{th} \cdot C_{th}} \qquad n = 1 \ldots \infty \quad (4)$$

$$p_1 = 375 \rightarrow t_1 = 2.664 \text{ ms}$$
$$p_2 = 3377 \rightarrow t_2 = 0.296 \text{ ms} \quad (5)$$

The corresponding magnitudes to these time constants are:

$$R_{Fn} = R_{th} \frac{8(-1)^{n+1}}{\pi^2 (2n-1)^2} \cdot \sin\left(\frac{\pi}{2}(2n-1)\frac{x}{L}\right) \quad (6)$$

$$R_{F1} = 45800\ \Omega \rightarrow C_{F1} = t_1 / R_{F1} = 58.2 \text{ nF}$$
$$R_{F2} = 5050\ \Omega \rightarrow C_{F2} = t_2 / R_{F2} = 58.6 \text{ nF} \quad (7)$$

Now the Foster model will be transformed to the Cauer one.

$$R_{C1} = 18310\ \Omega \qquad C_{C1} = 29.2 \text{ nF}$$
$$R_{C2} = 32550\ \Omega \qquad C_{C2} = 44.6 \text{ nF} \quad (8)$$

The hence generated R-C ladder, shown in Figure 2, has a true physical sense for the cantilever with lumped thermal resistors and capacitances.

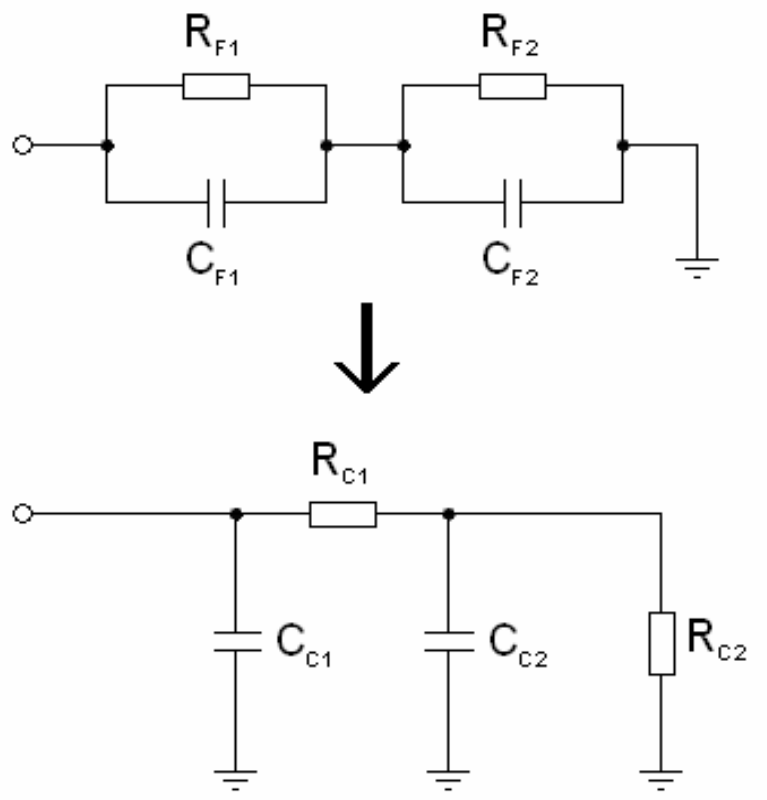

Figure 2. Foster to Cauer parameter matching

B. *Verifying the functionality*

In our thermal functional circuit the heating resistor generates a power equal of the square of the input voltage divided by its resistance. If we multiply it with the thermal impedance of the system we have the temperature difference between the cold and hot points of the thermopiles. Taking into account the Seebeck-effect, the output voltage is obtained.

Before complex investigations, this quadratic behaviour has to be proved in order to verify the operability of the microsystem. For this purpose, the electro-thermal simulation algorithm presented in [5] was used. For cross-verification, some measurements were performed as well, to confirm the results. Two types of measurement and simulation have been performed. First the heating resistor was driven by DC input which resulted in smaller amplitude DC output at the terminals of the thermopiles. The results are presented in Figure 3, where the scaling of the amplitude and quadratic dependence can be seen.

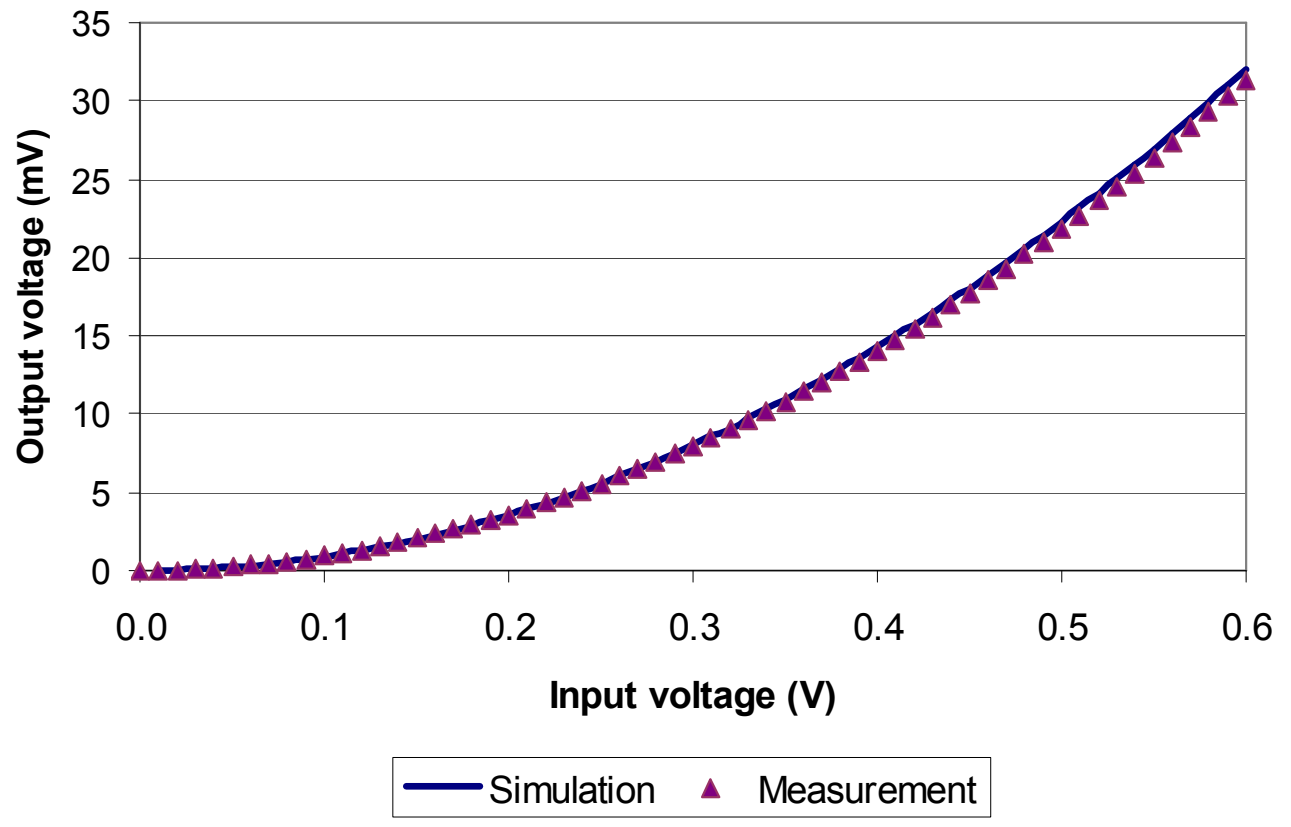

Figure 3. DC transfer characteristic

The measured and simulated amplitudes indicate good correspondence to each other and their maximum difference is 0.732 mV at 0.6 V input voltage which can be observed in Figure 4.

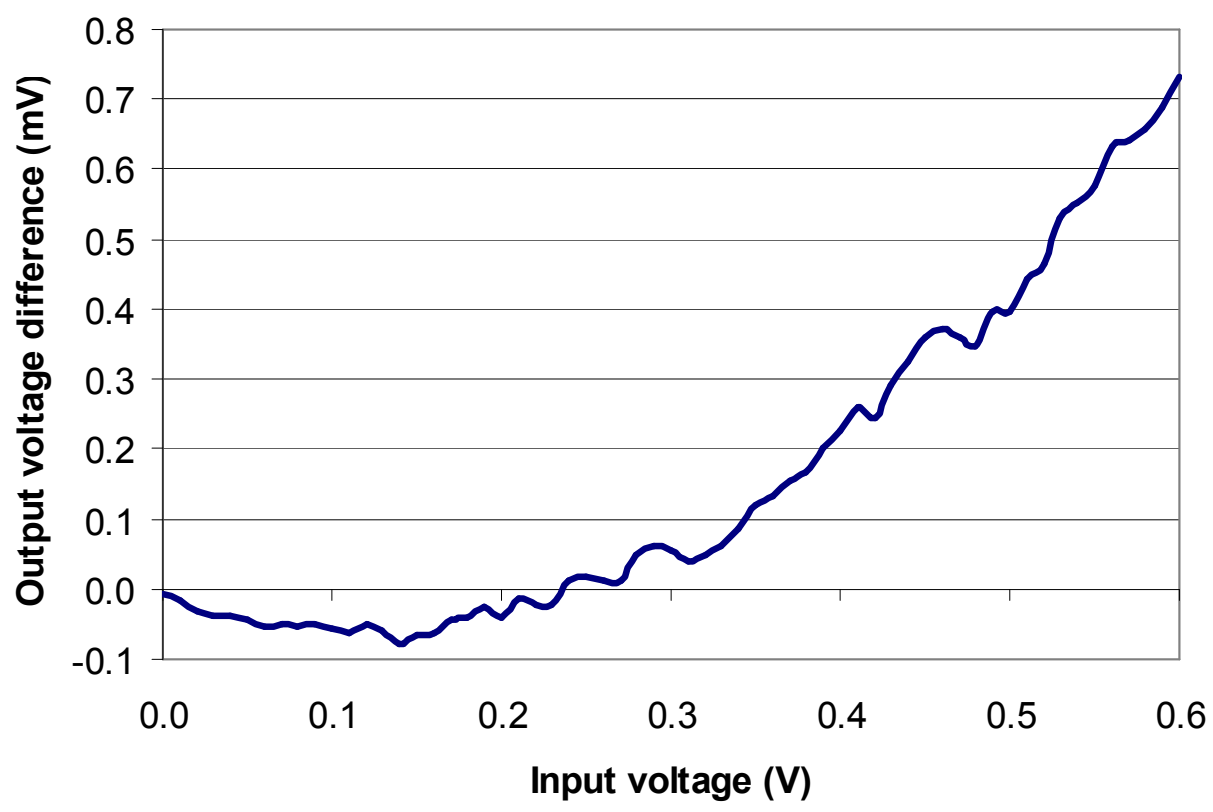

Figure 4. Difference between the simulated and the measured voltage amplitudes

From these characteristics we can define a $K$ sensitivity or conversion constant which shows that how the output voltage amplitude is determined by the square of the input amplitude, namely

$$K = \frac{U_{out}}{U_{in}^2} \cdot \qquad (9)$$

Reading the simulated and measured functions from Figure 3, we obtain

$$K_{meas} = 0.0857 \frac{1}{V} \qquad (10)$$

for the measured sensitivity and

$$K_{sim} = 0.0893 \frac{1}{V} \qquad (11)$$

for the simulated conversion constant.

In the other measurement the resistor was driven by a sinusoidal time-domain input. Obviously the time vs. output voltage plots in Figure 5 show frequency doubling effect with a minor offset.

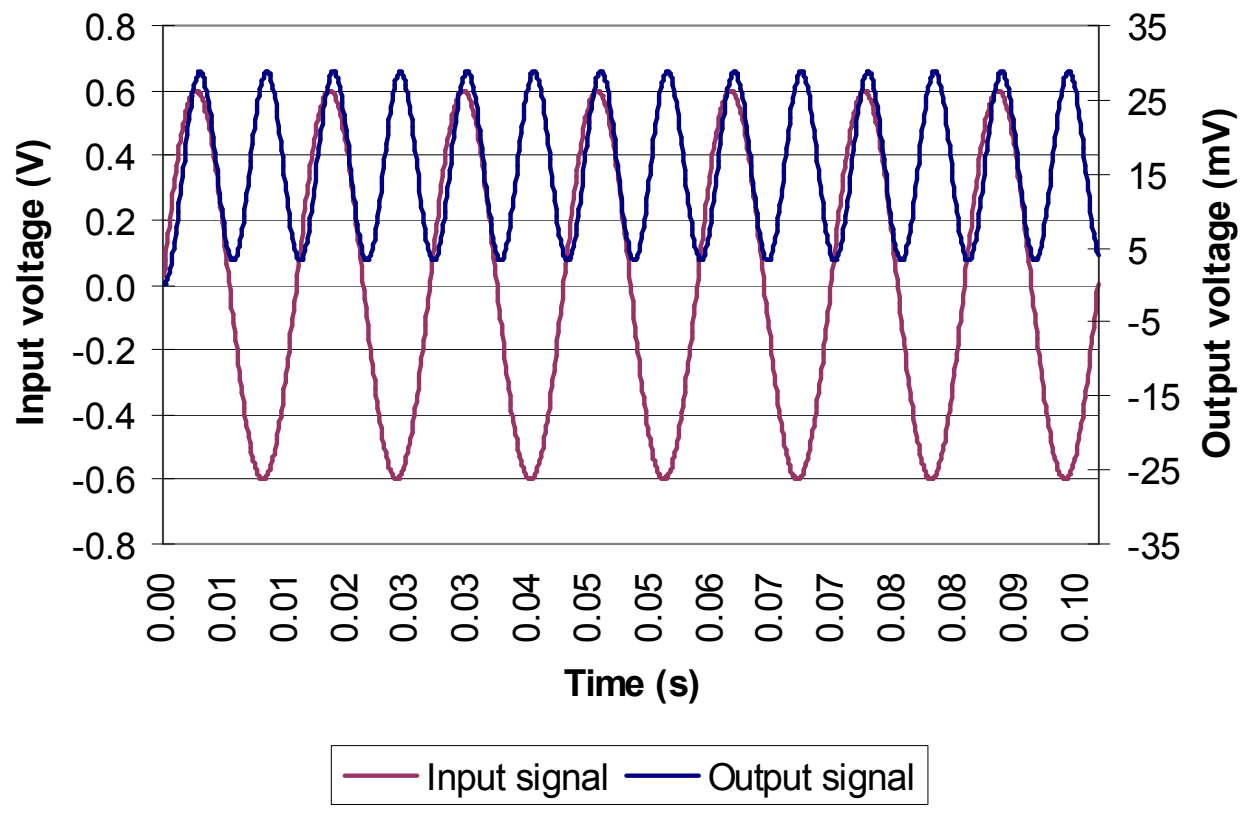

Figure 5. Time vs. voltage plot

## III. TEMPERATURE DEPENDENCIES

In the next phase of the characterization the temperature dependency of the conversion constant ($K$) was measured. There are three different factors, which has thermal dependency, namely the thermal conductivity ($\lambda$), the Seebeck-coefficient ($S$), and the resistance of the heating resistor ($R_H$). If their dependency is approximated linearly in room temperature, then their thermal behaviour can be described as

$$R_H = R_{H0} \cdot [1 + \alpha_R \cdot (T - T_0)] \qquad (12)$$

$$\lambda = \lambda_0 \cdot [1 - \alpha_\lambda \cdot (T - T_0)] \qquad (13)$$

$$S = S_0 \cdot [1 - \alpha_S \cdot (T - T_0)] \qquad (14)$$

$$K \cong K_0 \cdot [1 + \alpha_K \cdot (T - T_0)] \qquad (15)$$

where the *0* indexed members are the magnitudes at $T_0$ temperature and the $\alpha$ members are the thermal coefficients which are in connection with each other by

$$\alpha_K = \alpha_\lambda - \alpha_R - \alpha_S \, . \qquad (16)$$

Usually $\alpha_K$ is larger than zero which means that the conversion efficiency rises with the increase of the temperature.

To prove this theory a measurement was set up where the temperature change in the device was achieved by external thermal excitation. At the end of the first measurement negative prefix result was occurred, which means that the conversion constant is decreasing with increasing temperature.

In order to verify this behaviour the components of the conversion constant are measured and mapped. The most obvious was to measure the change of the impedance of the resistor. Figure 6 shows this change in the range 10-90 °C.

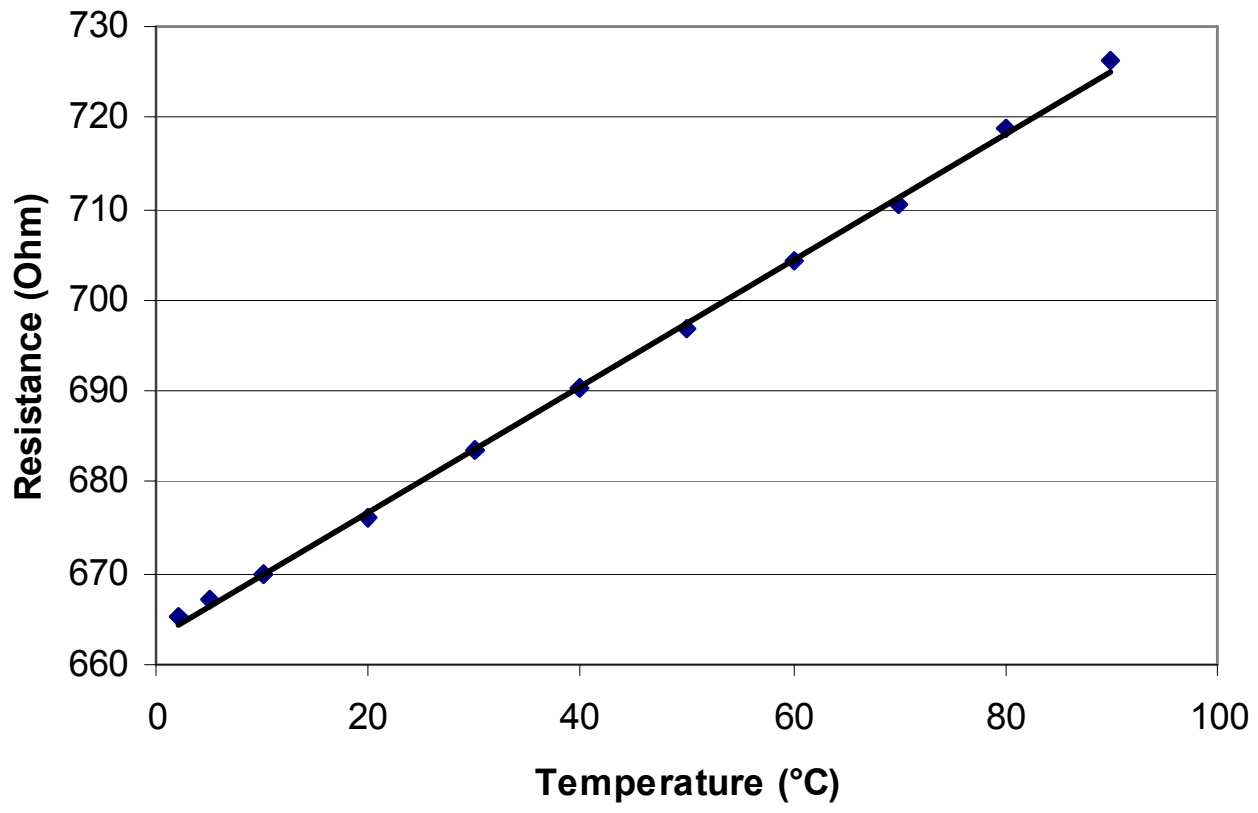

Figure 6. Temperature dependency of the heating resistor

Calculating its thermal coefficient we got:

$$\alpha_R = \frac{\frac{R_f}{R_{f0}} - 1}{(T - T_0)} = \frac{\frac{726.18}{670.01} - 1}{90 - 10} = 0.00105 \frac{1}{\mathrm{K}} \qquad (17)$$

The resultant thermal conductivitiy coefficient of the system was calculated using area compensation like (1) and

$$\alpha_\lambda = \frac{\sum \alpha_{\lambda i} \cdot d_i}{\sum d_i} = 0{,}00177 \frac{1}{\mathrm{K}} \qquad (18)$$

was received, where the coefficients of the polysilicon and $SiO_2$ were taken from [6] and [7].

Inserted it into (16) we got a 0.00113 for the Seebeck-constant thermal dependency, which is in agreement with the values from [8].

## IV. NON-LINEAR AND DISTORTION ANALISYS

In the last chapter we examined the non-linearity and the distortion in the transfer function. One way to determine this characteristic is to measure the upper harmonics in the response signal of the system. If we investigate the peaks in the spectrum, it can reveal which parameter or part of the structure contains additional elements which cause non-linearity. Here, the DC and the two omega components are evident because of the quadrate function. As the thermal coefficients of Seebeck-effect, the heat conductivity and the heating resistors are temperature dependent, this can cause the appearance of the higher order harmonics.

### *A. Spectrum analysis*

Since the output amplitude is in mV range and the magnitude of the unwanted harmonics are even smaller, the shielding and the use of appropriate devices subject of the spectrum analysis are more important than before. In order to have a relatively high resolution output function a computer aided measurement which can be seen in Figure 7 was used, where the output signal was connected to the sound card of a computer where it was recorded and evaluated. But afore this step, the amplitude has to be amplified for the card in a way which will not provide upper harmonics. For this purpose a linear amplifier was constructed. Although this circuit contains low pass elements, they don't bother us because their cut-off frequency is much higher than the investigated frequency range.

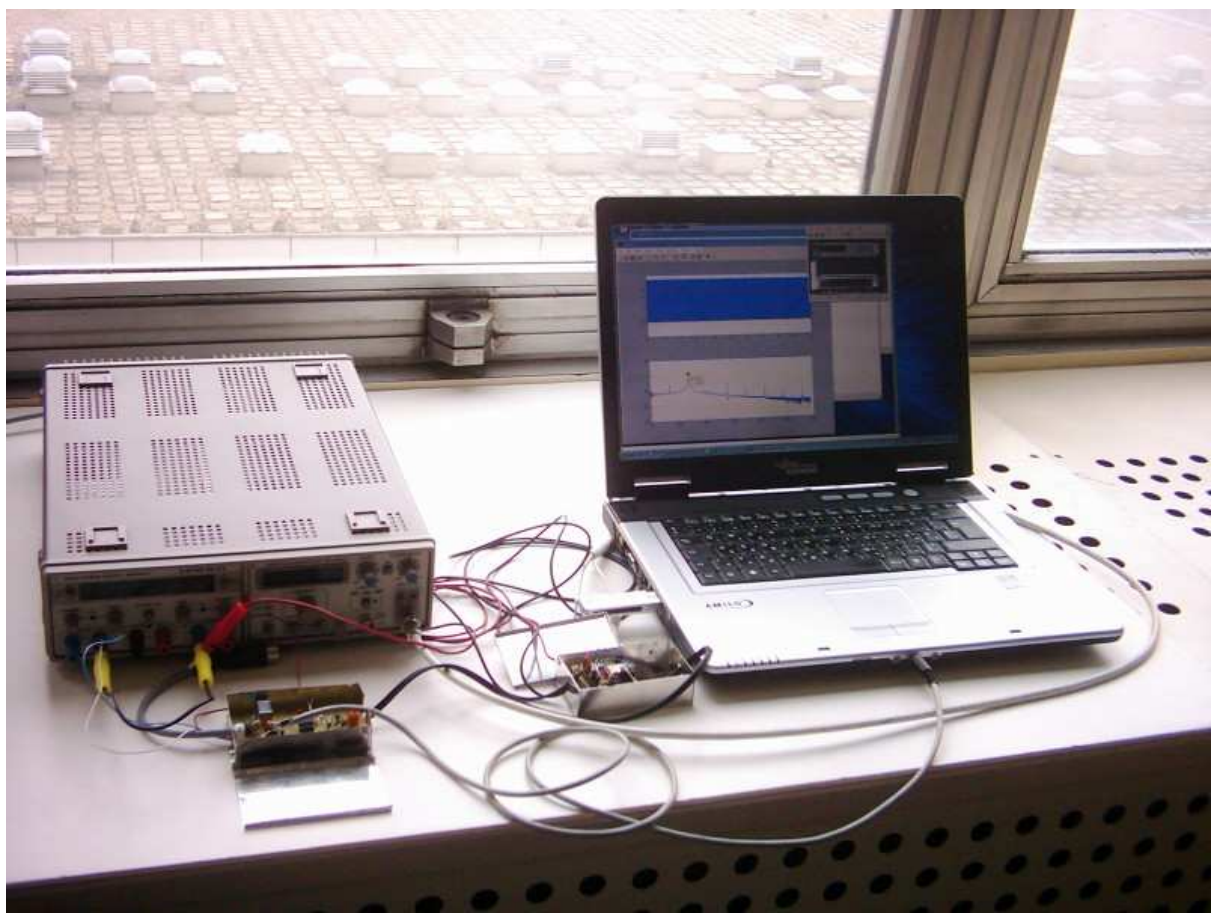

Figure 7. Measurement setup

During the analyses of the recorded file three types of high resolution windowing function were used, namely Hamming, Hann and Bartlett, in order to have comparable results and to filter out the outer distortion as much as we can. Out of these three, the Hamming windowed spectrum is presented in Figure 8. All the calculated frequency spectrums show the peak at the second harmonic of the input signal due to the frequency doubling as it was expected.

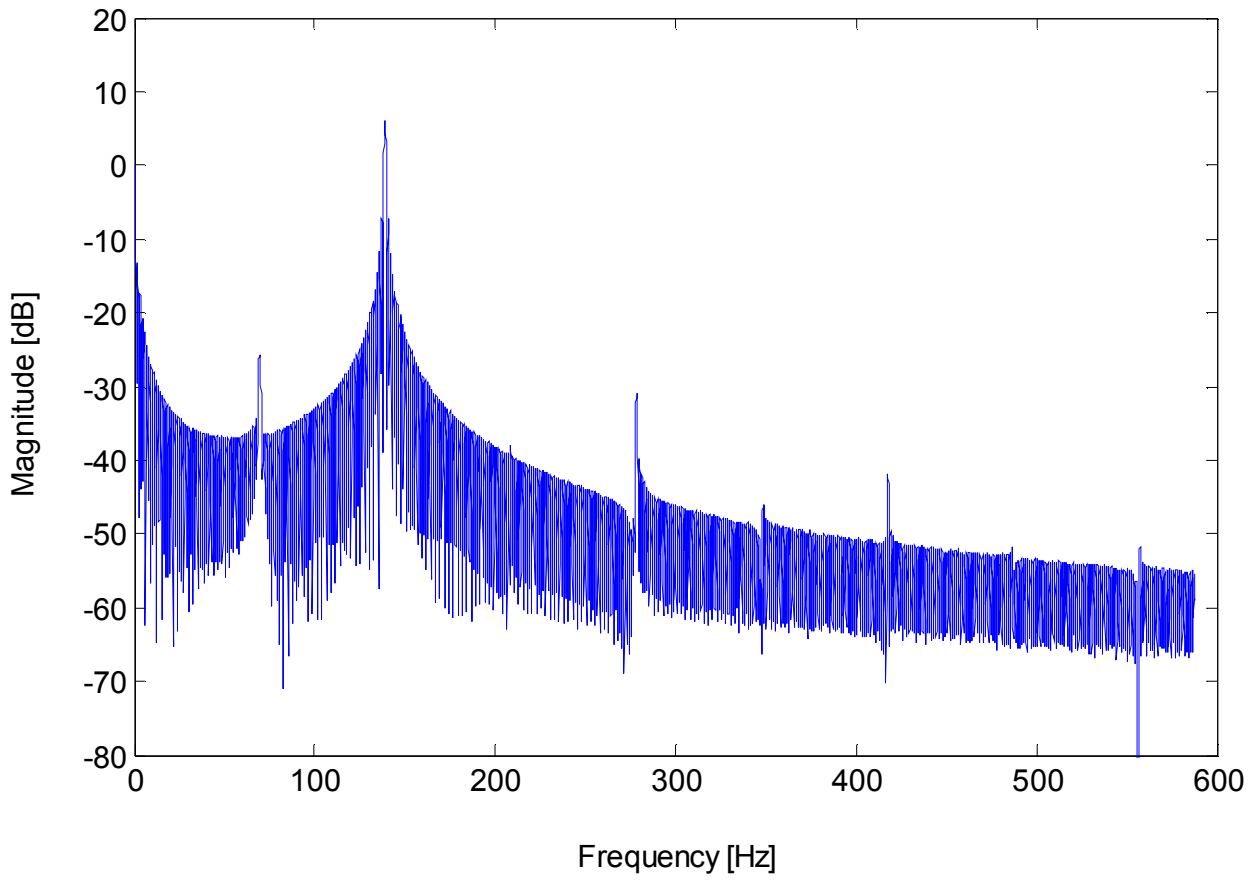

Figure 8. Magnitude vs. frequency plot ($f_1$=70 Hz)

The three characteristics also indicate additional integral upper harmonics and other components which are the result of the non-linear transfer characteristic of the device together with the distortion of the system. By decreasing the input amplitude, the magnitude of these upper harmonics decreased as well. Table 1 shows this correspondence between the input voltage and the non-linearity.

TABLE1
HARMONICS MAGNITUDE DEPENDENCE

| Frequency component | Magnitude at 1.25 V | Magnitude at 1.1 V | Magnitude at 0.95 V |
|---|---|---|---|
| DC | 0 dB | 0 dB | 0 dB |
| First harmonic (70 Hz) | -27.6 dB | -28.5 dB | -31.2 dB |
| Second harmonic (140 Hz) | 4.2 dB | 3 dB | 0.85 dB |
| Fourth harmonic (280 Hz) | -35.4 dB | -36 dB | -41.3 dB |
| Sixth harmonic (420 Hz) | -48.1 dB | -49.5 dB | -53.8 dB |

*B. Equivalent model generation*

By using the results of the calculations and the simulations, all the secondary parameters were determined and due to their consequence, information can be obtained about the distortion and the spurious components. Our further task was to construct an electrical model of the MEMS structure suitable to use in Spice like circuit simulation programs. Therefore we set up the electro-thermal equivalent circuit of the system which is shown in Figure 9.

The resistor at the left side represents the heating resistor, which generates

$$U_{in}^2 / R_H \qquad (19)$$

power. If we consider this power to be a thermal current, than we can represent it by a voltage controlled current source. The Cauer R-C ladder, connected to this source represents thermal behaviour of the cantilever with the previously calculated values. Here we modeled the temperatures by electronic potentials and therefore we said that the $T_0$ temperature of substrate is replaced with the electrical ground and the $T_H$ temperature which is at the hot spot of the thermopiles with $U_H$. After this matching, the thermo-electrical transformation based on the Seebeck-effect, can be considered as a voltage-driven voltage source with a gain of

$$U_S = N \cdot S \cdot U_H \qquad (20)$$

where $N$ is the number of serially connected thermocouples. Next to it the overall resistance of thermopiles is presented.

Although this model describes the quadrate function, it doesn't give us information about the other harmonics. Therefore we had to consider using variable model elements.

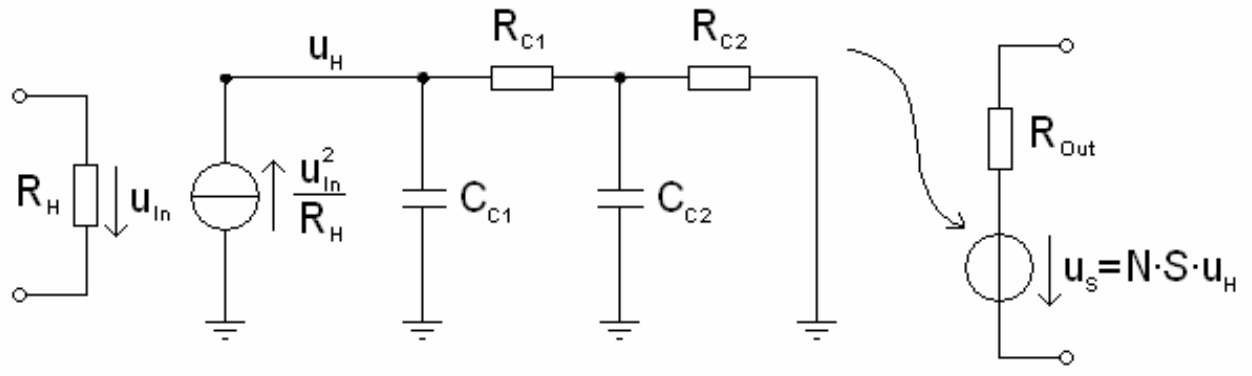

Figure 9. Electro-thermal equivalent circuit

The fundamental frequency which is 28 dB lower than the doubled frequency comes from the connecting wires because the input and output lines are so close to each other that coupling can be occurred. It is represented by a $+U_{in}$ component in the $U_S$. The odd harmonics are the consequence of the temperature change. As for the generated temperature difference, $T_H - T_0$, is matched with $U_H$, and therefore, all the temperature dependent elements have a voltage controlled component in the model. One such thing is the heating resistor. Figure 4 shows its linear temperature dependency wherefore it can be calculated as

$$R_H = R_{H0} \cdot (1 + \alpha_R \cdot U_H) \qquad (21)$$

The temperature dependency of the thermal conductivity brings a

$$\frac{1}{1 - \alpha_\lambda \cdot U_H} \qquad (22)$$

multiplier to the distributed $R_{th}$ calculation, where $\alpha_\lambda$ is the thermal conductivity's thermal coefficient, which causes the same multiplication in each of the Cauer R-C ladder resistors.

As discussed before the Seebeck-effect driven thermo-electrical transformation also has a temperature dependent component, which result

$$U_S = S \cdot U_H \cdot (1 - \alpha_S \cdot U_H) \qquad (23)$$

where $\alpha_S$ is the thermal coefficient of the Seebeck-effect. As all the dependencies were taken into account, our model was modified. The new model is presented in Figure 10.

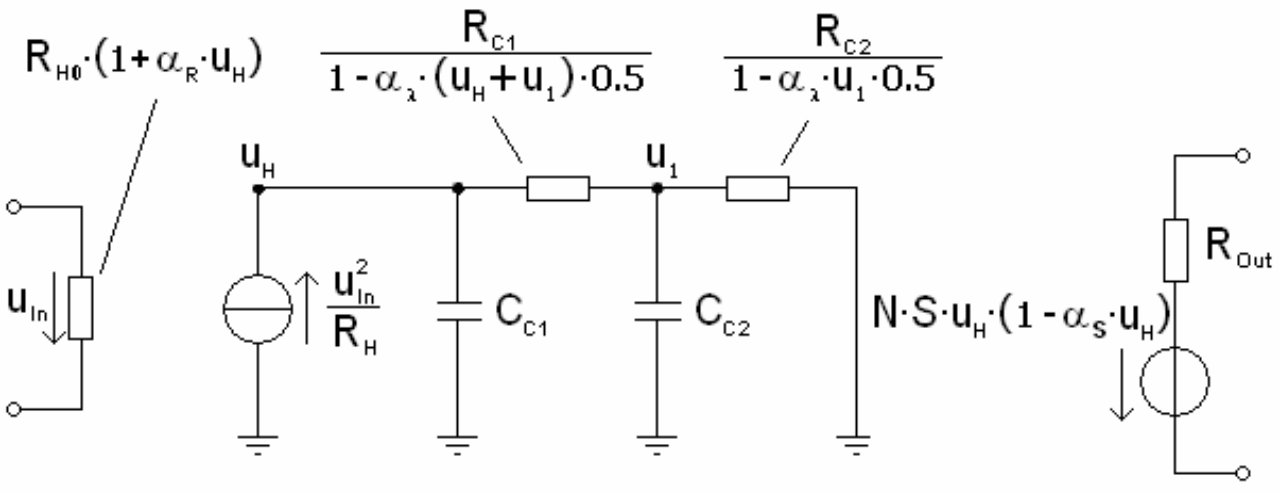

Figure 10. Modified electro-thermal equivalent circuit

With the calculation of the dependencies our model describes the non-linear behaviour of the RMS meter except some minor peaks which amplitude is about 50 dB lower than the one of the DC component so they are considered to be minor distortions. In addition we performed transient analyses with our model in a SPICE based simulator and the results were in good agreement with the measured characteristic.

## V. Conclusions

The presented case study shows a characterization process where we investigated an electro-thermal test-MEMS and through this, some statements about the properties of the structure were realized. The article described how the parameters react to the piled heat amount and their manifestation in different applications. Through these observations further research can be made in order to advise new design issues to minimize the distortions and the unwanted non-linearity.

## References

[1] Vladimír Székely: "New type of thermal-function IC: The 4-quadrant multiplier", *Electronics Letters*, Vol. 12, No. 15, pp. 372-373., 22nd July 1976

[2] Martin von Arx, Oliver Paul, Henry Baltes, "Process-dependent thin-film thermal conductivities for thermal CMOS MEMS", IEEE *Journal of Microelectromechanical Systems*, Vol. 9, No. 1, March 2000

[3] Vladimír Székely, "THERMODEL: A tool for compact dynamic thermal model generation", *Elsevier, Microelectronics Journal* 29, pp. 257-267., 1998

[4] Vladimír Székely, Márta Rencz, András Poppe, Bernard Courtois, "THERMODEL: A tool for thermal model generation and application for MEMS", *Springer*, *Journal Analog integrated circuits and signal processing* 29, pp. 49-59., 2001

[5] Márta Rencz, Vladimír Székely, András Poppe, Bernard Courtois, "Electro-thermal simulation of MEMS elements", *IEEE, DTIP 2003 of MEMS & MOEMS*

[6] P.D. Maycock: "Thermal conductivity of silicon, germanium, III-V compounds and III-V alloys", *Solid-state electronics*, Vol. 10, pp. 161-168., 1967

[7] Arokia Nathan, Henry Baltes: "Microtransducer CAD: Physical and Computational Aspects", Springer, 1999

[8] T. H. Geballe G.W. Hall: "Seebeck effect in silicon", *Physical Review*, Vol. 98, No. 4. pp. 940-947., 1955